# Power-Efficient Resource Allocation in Massive MIMO Aided Cloud RANs

Nahid Amani
Department of Communication Technologies
Iran Telecommunication Center,
Tehran, Iran
n_amani@itrc.ac.ir

Saeedeh Parsaeefard
Department of Communication Technologies
Iran Telecommunication Center,
Tehran, Iran
s.parsaeifard@itrc.ac.ir

Hassan Taheri
Department of Electrical Engineering
AmirKabir University of Technology Tehran, Iran
htaheri@aut.ac.ir

Hossein Pedram
Department of Computer Engineering
AmirKabir University of Technology Tehran, Iran,
pedram@aut.ac.ir

*Abstract*—**This paper considers the power-efficient resource allocation problem in a cloud radio access network (C-RAN). The C-RAN architecture consists of a set of base-band units (BBUs) which are connected to a set of radio remote heads (RRHs) equipped with massive multiple input multiple output (MIMO), via fronthaul links with limited capacity. We formulate the power-efficient optimization problem in C-RANs as a joint resource allocation problem in order to jointly allocate the RRH and transmit power to each user, and fronthaul link and BBU assign to active RRHs while satisfying the minimum required rate of each user. To solve this non-convex optimization problem we suggest iterative algorithm with two-step based on the complementary geometric programming (CGP) and the successive convex approximation (SCA). The simulation results indicate that our proposed scheme can significantly reduce the total transmission power by switching off the under-utilized RRHs.**

## I. Introduction

With the explosive growth of mobile data traffic, the fifth-generation (5G) wireless networks encounter considerable challenges in enhancing spectrum efficiency (SE) and energy efficiency (EE). Cloud radio access network (C-RAN) and massive multiple input multiple output (MIMO) are two promising approaches to tackle these challenges for 5G [1-3].

Massive MIMO base stations (BSs) will be able to communicate with multiple single-antenna users over the same time-frequency slot, therefore, by providing a high power gain, decreases the transmit power which leads to improve SE and EE [4, 5]. Also, C-RAN as a novel RAN architecture separates remote radio head (RRH) from base band units (BBUs), which helps to provide cost efficient and flexible deployment of traditional base stations (BSs).

Hence, by deploying a large number of RRHs in a cell, with less transmission power of each RRH, SE and EE will be significantly improved. Additionally, equipping each RRH with many antennas (massive MIMO), can lead to higher SE and EE to 5G [6- 9]. This scales up the complexity gain of traditional MIMOs [1], [10] and reduces the interference among users of all access points [1].

Dense deployment of RRHs to reach higher SE causes under-utilized RRHs which consequently leads to increasing energy consumption [11]. Additionally, due to existence of interference among users in different highly overlapped areas of RRHs, user assignment to RRHs based on the signal strength is insufficient. Therefore, to reduce interference and enhance power efficiency, power control becomes challenging issue. To address these challenges, we formulate the power-efficient optimization problem with the novel utility function in order to minimize the operation cost. Hence, we consider the summation of energy consumption cost of RRHs and total transmit power of all users as operation cost in utility function.

In recent years a large number of studies explored the optimization of SE and EE in massive MIMO and C-RANs. In [12], by optimally assigning users, beamforming parameters are adjusted in order to maximizing EE in C-RAN. In this paper, by using norm approximation, non-convex formulated optimization problem is transformed into the convex one. [13] Formulates a power minimization beamforming problem in C-RAN. With switching off RRHs, the power consumption of RRHs and fronthaul links are reduced by applying a group sparse beamforming method, regardless capacity limitation of fronthaul links. [14] reduces the overall power consumption by employing an online stochastic game theoretic to learn the cellular traffic patterns and switch on-off the RRHs. [15] formulates an EE optimization problem, and uses Lagrange dual decomposition method to minimize the overall power consumption with access point (AP) and power allocation to each user in H-CRAN. [11] reduces the power consumption in C-RAN by switching off RRHs and proposes a heuristic algorithm to solve the RRH-BBU assignment and allocates RRH to each user, regardless capacity limitation of BBUs and fronthaull links. [16-18] investigate optimal power and range adaptation policies with time-varying traffic to minimize the average power consumption of APs. Interference reduces by cooperating between different Aps which leads to switch off the APs with low data traffic and decreases the overall power consumption. [7] formulates joint sub-carrier, power, AP and activated antennas allocation with the aim of minimizing the total power consumption and maximizing SE simultaneously in massive MIMO enabled heterogeneous networks (HetNets). The authors of [19] by employing a stochastic geometry method, analyze that using flexible cell association can improve the EE of HetNets by offloading data traffic to small cell in massive MIMO enabled HetNets. The authors of [20] employ both the cooperative and noncooperative EE power control game in multi-tier MIMO HetNets, where all tiers cooperatively or selfishly choose their transmit power to maximize their network EE.

None of the mentioned works has jointly considered resource allocation and C-RAN limitations such as maximum load capacity of BBUs to minimize the total network transmission power by switching off under-utilized RRHs in MIMO aided C-RAN. To address this gap, the aim of this paper is to study the joint optimization problem with the transmit power and AP allocation subject to fronthaul and RRH assignment to BBU and interference mitigation to minimize the total network transmission power by switching off under-utilized RRHs in MIMO aided C-RAN system. Hence, regarding the variations of traffic load, the RRHs with low data traffic (under-utilized) and their corresponding fronthaul links will be switched off according to the minimum power consumption cost of RRHs. Therefore associated users to them are moved to the neighboring RRHs.

Due to interference among users from various RRHs and existence of integer variables such as RRH assignment to each user and RRH allocation to BBUs, the formulated optimization problem is non-convex and NP-hard with high computational complexity [21]. We apply the complementary geometric programming (CGP) and the successive convex approximation (SCA) [22-26] to develop a two-step iterative algorithm with low computational complexity to solve the formulated problem. Via different relaxation and transformation techniques such as DC-approximation and arithmetic-geometric mean approximation (AGMA), the sub-problems in each step are converted into its geometric programming (GP) problem [27, 28], which will be solved via optimization software packages like CVX [29].

We compare performance of the proposed approach with traditional approach where each user is associated to the RRHs based on largest value of received SINR. The simulation results reveal that our proposed approach with novel utility function is more efficient than the traditional approach, in terms of increasing total EE and reducing total network transmission power. The simulation results show that total transmission power is reduced more than 20% compared to that of the traditional algorithm for dense region. Also, simulation results illustrate that our proposed algorithm can effectively move associated users from under-utilized RRHs to neighboring RRHs which lead to switch off under-utilized RRHs.

In the remainder of this paper, Section II presents the system model and formulated optimization problem to minimize the total network transmission power. Section III introduces the two-step iterative algorithm. Section IV indicates the simulation results, followed by concluding remarks in Section V.

II. SYSTEM MODEL AND PROBLEM FORMULATION

Consider a downlink transmission in a C-RAN architecture that serves a set of $\mathcal{N} = \{1,...,N\}$ single-antenna users by a set of $\mathcal{R} = \{1,...,R\}$ RRHs as shown in Fig. 1. In this specific region, each RRH $r \in \mathcal{R}$ is equipped with $F_r \gg 1$ antennas and connected to a set of $\mathcal{B} = \{1,...,B\}$ limited-capacity BBUs via a fronthaul link. BBUs are responsible to process the baseband signals. We define $\alpha_{r,n}$ for association between user $n \in \mathcal{N}$ and RRH $r \in \mathcal{R}$ as

$$\alpha_{r,n} = \begin{cases} 1, & \text{if RRH } r \text{ is associated to } n^{th} \text{ user}, \\ 0, & \text{otherwise}. \end{cases}$$

Suppose $p_{r,n}$ be the transmit power and $h_{r,n}$ be channel gain from the RRH $r \in \mathcal{R}$ to the user $n \in \mathcal{N}$, also the number of transmit antennas $F_r$ be much more than the number of simultaneously served users by a RRH $r$. Under these assumptions, according to [7, 30], the achievable rate (throughput) of each user $n$ at RRH $r$ can be expressed as

$$R_{r,n}(\mathbf{P},\boldsymbol{\alpha}) = \log_2(1+(\frac{F_r - N_r + 1}{N_r}\frac{p_{r,n}h_{r,n}}{\sigma^2 + I_{r,n}})), \quad (1)$$

In which $N_r = \sum_{n \in \mathcal{N}} \alpha_{r,n}, \forall r \in \mathcal{R}$ represents the total number of users allocated to the RRH $r$ and $I_{r,n} = \sum_{\forall r' \in \mathcal{R}, r' \neq r} \sum_{\forall n' \neq n} p_{r',n'}h_{r,n'}$, denotes the interference to user $n \in \mathcal{N}$ in RRH $r \in \mathcal{R}$, and $\sigma^2$ is the noise power which is assumed to be equal for all users.

To save energy, a RRH and corresponding fronthaul links can be switched off, when traffic of associated users to them is low. Therefore, $y_r$ is defined for the on-off states of RRH $r \in \mathcal{R}$ as

$$y_r = \begin{cases} 1, & \text{if RRH r is in state on,} \\ 0, & \text{otherwise,} \end{cases}$$

thus, connection between RRH $r \in \mathcal{R}$ and BBU $b \in \mathcal{B}$ is defined as

$$\beta_{r,b} = \begin{cases} 1, & \text{if the RRH r is assigned to the BBU b,} \\ 0, & \text{Otherwise.} \end{cases}$$

Furthermore, $\mathbf{P}$, $\boldsymbol{\alpha}$, $\mathbf{Y}$, and $\boldsymbol{\beta}$ are matrices of all $p_{r,n}, \alpha_{r,n}, y_r$ and $\beta_{r,b}$, respectively, for all $n \in \mathcal{N}, r \in \mathcal{R}$ and $b \in \mathcal{B}$.

With the aim of reducing network energy consumption cost, we define a novel network utility function as

$$U(\boldsymbol{\alpha},\mathbf{P},\mathbf{Y}) = \sum_{r \in R}\sum_{n \in N} \alpha_{r,n}p_{r,n} + C_a \sum_{r \in R} y_r F_r, \quad (2)$$

which is summation of total transmit power of all users and the energy consumption cost of active RRHs. The parameter $C_a$ is proportional to transmit power of each antenna in active RRHs. The simultaneous reduction of these two costs is the novelty of dynamic resource allocation in this work. Consequently, based on (2), the optimization problem to minimize the total energy consumption cost can be expressed as

$$\min_{\boldsymbol{\alpha},\boldsymbol{\beta},\mathbf{Y},\mathbf{P}} U(\boldsymbol{\alpha},\mathbf{P},\mathbf{Y}), \quad (3)$$

subject to:

$C1: \sum_{n \in \mathcal{N}} p_{r,n} \leq p_r^{\max}, \forall r \in \mathcal{R},$

$C2: \sum_{r \in \mathcal{R}} \alpha_{r,n} R_{r,n}(\mathbf{P},\boldsymbol{\alpha}) \geq R_n^{\text{rsv}}, \forall n \in \mathcal{N},$

$C3: \sum_{r \in R} \alpha_{r,n} \leq 1, \forall n \in \mathcal{N},$

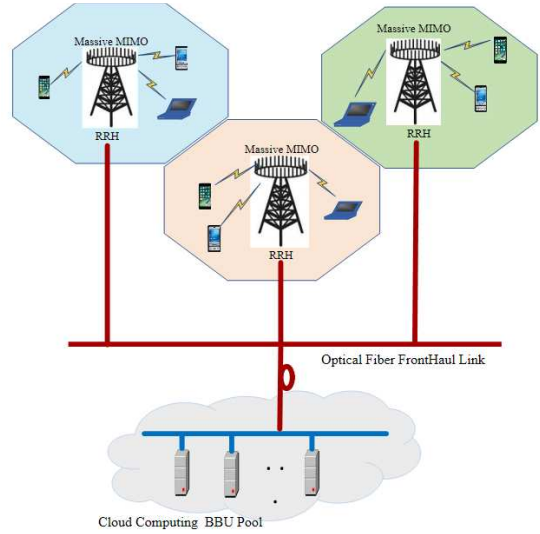

Fig. 1. C-RAN architecture with cloud computing BBU pool and massive MIMO RRHs.

$C4: \sum_{b \in \mathcal{B}} \beta_{r,b} \leq 1, \forall r \in \mathcal{R},$

$C5: \sum_{r \in \mathcal{R}}\sum_{n \in \mathcal{N}} \beta_{r,b}\alpha_{r,n} R_{r,n}(\mathbf{P},\boldsymbol{\alpha}) \leq L_b^{\max}, \forall b \in \mathcal{B}$

$C6: \sum_{b \in \mathcal{B}} \beta_{r,b} - y_r \leq 0, \forall r \in \mathcal{R},$

$C7: \sum_{n \in \mathcal{N}} \alpha_{r,n} \leq y_r \times \omega, \forall r \in \mathcal{R},$

$\alpha_{r,n} \in \{0,1\}, y_r \in \{0,1\}, \beta_{r,b} \in \{0,1\}, \forall r,n,b.$

In (3), C1 indicates that maximum transmit power of each RRH $r \in \mathcal{R}$ is restricted by $p_r^{\max}$. C2 denotes the minimum required rate, i.e., $R_n^{\text{rsv}}$, for each user. C3 represents that each user can be only associated to at most one RRH at any time. C4 denotes each RRH $r \in \mathcal{R}$ can be associated to at most one BBU. The allocated load to each BBU is received from its own corresponding associated RRHs. Therefore, C5 specifies that the maximum load supported by each BBU $b \in \mathcal{B}$ is restricted by $L_b^{\max}$. Based on C6, when RRH $r \in \mathcal{R}$ is in state on its corresponding fronthaul link can be enabled. Finally, C7 indicates each user $n \in \mathcal{N}$ can be assigned to RRH $r$ when RRH $r$ is in state on and $\omega$ is constant value.

Due to the interference term in C2 and some integer variables such as $\alpha_{r,n}, \beta_{r,b}, y_r$, the formulated problem (3) is non-convex and NP hard with high computational complexity [22]. To overcome this issue, we propose the two-step iterative solution algorithm with low computational complexity to solve the formulated optimization problem by employing the SCA and CGP. In the next section, we explain our algorithm to solve (3).

## III. TWO-STEP ITERATIVE ALGORITHM FOR JOINT CLOUD PARAMETERS ASSIGNMENT AND POWER ALLOCATION

To obtain an efficient solution for (3), a two-step iterative solution algorithm is developed with the aim to separate the transmission and cloud parameters. Step 1 specifies the cloud parameters for joint RRH, fronthaul and BBU allocation, while the Step 2 achieves the transmission parameter, consisting the transmit power. Therefore, in Step 1, with a given (fixed) power allocation vector, the optimal $\boldsymbol{\alpha}, \boldsymbol{\beta}$ and $\mathbf{Y}$ vectors are derived. Then, based on the values obtained from Step 1, transmit power is allocated to each user in Step 2. The whole process is expressed as follows

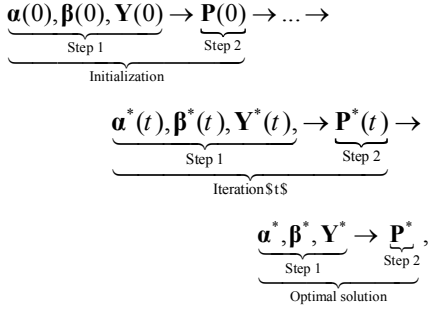

where $t \geq 0$ is the iteration index. Also, $\boldsymbol{\alpha}^*(t), \boldsymbol{\beta}^*(t), \mathbf{Y}^*(t)$ and $\mathbf{P}^*(t)$ are optimal values obtained at iteration $t$. The iterative procedure is stopped when convergence criteria are met, e.g.,

$$||\boldsymbol{\alpha}^*(t) - \boldsymbol{\alpha}^*(t-1)|| \leq \epsilon_1, ||\boldsymbol{\beta}^*(t) - \boldsymbol{\beta}^*(t-1)|| \leq \epsilon_2,$$
$$||\mathbf{Y}^*(t) - \mathbf{Y}^*(t-1)|| \leq \epsilon_3, ||\mathbf{P}^*(t) - \mathbf{P}^*(t-1)|| \leq \epsilon_4,$$

where $0 < \epsilon_1, \epsilon_2, \epsilon_3, \epsilon_4 \ll 1$ Note that, the sub-problems of Steps 1 and 2 are still non-convex and encounter high computational complexity. To solve them, we first relax the integer variables then, by applying various transformation and DC-approximation in propositions 1, 2, and 3 we try to transform the non-convex sub-problems in Steps 1 and 2 into the equivalent lower-bound standard form of GP. To get more about CGP, refer to Section III.A in [22].

*Step 1: Cloud Parameters Allocation Algorithm*

In each iteration $t$, this step drives optimal values of $\boldsymbol{\alpha}, \boldsymbol{\beta}$ and $\mathbf{Y}$ with fixed values of $\mathbf{P}(t)$. Hence, (3) is simplified into

$$\min_{\boldsymbol{\alpha}, \boldsymbol{\beta}, \mathbf{Y}} U(\boldsymbol{\alpha}, \mathbf{P}(t), \mathbf{Y}) \tag{4}$$

subject to: $C2 - C7,$

Where in (4), optimization variables are $\boldsymbol{\alpha}, \boldsymbol{\beta}$ and $\mathbf{Y}$, therefore, (4) has less computationally complex than (3).

Since we assumed that the number of transmit antennas $F_r$ be much more than the number of simultaneously served users by a RRH $r \in \mathcal{R}$ as $F_r \gg N_r$, from this assumption we can write $\frac{F_r - N_r(t_1) + 1}{N_r(t_1)} \approx \frac{F_r}{N_r(t_1)}$. Hence, with considering fixed value of $\mathbf{P}(t)$ and high SINR scenario, we can rewrite (1) as

$$\tilde{R}_{r,n}(\mathbf{P}, \boldsymbol{\alpha}) \approx \log_2(\frac{F_r}{N_r(t_1)} \gamma_{r,n}(t)), \forall r \in \mathcal{R}, \tag{5}$$

In which $\gamma_{r,n}(t) = \frac{p_{r,n}(t) h_{r,n}}{\sigma^2 + \sum_{r' \neq r} \sum_{n' \neq n} p_{r',n'}(t) h_{r,n'}}$ is signal-to-interference-plus noise ratio (SINR) of user $n \in \mathcal{N}$ at RRH $r \in \mathcal{R}$. Therefore, at the iteration $t_1$, (4) is converted into

$$\min_{\boldsymbol{\alpha}, \boldsymbol{\beta}, \mathbf{Y}} U(\boldsymbol{\alpha}, \mathbf{P}(t), \mathbf{Y}), \tag{6}$$

subject to: $C3, C4, C6, C7,$

$C2.1: \sum_{r \in \mathcal{R}} \alpha_{r,n} \tilde{R}_{r,n}(\mathbf{P}(t), \boldsymbol{\alpha}) \geq R_n^{\text{rsv}}, \forall n \in \mathcal{N},$

$C5.1: \sum_{r \in \mathcal{R}} \sum_{n \in \mathcal{N}} \beta_{r,b} \alpha_{r,n} \tilde{R}_{r,n}(\mathbf{P}(t), \boldsymbol{\alpha}) \leq L_b^{\max}, \forall b \in \mathcal{B}$

(6) is non-convex with high complexity due to the integer optimization variables, e.g., $\boldsymbol{\alpha}, \boldsymbol{\beta}, \mathbf{Y}$, and the non-convex constraints. C2.1, C5.1 and C6 are not in a GP standard form due to the logarithm term in the rate formula of C2.1 and C5.1, and negative terms in C6. To solve these issues, we first relax the integer variables as $\alpha_{r,n} \in [0,1], \beta_{r,b} \in [0,1]$ and $y_r \in [0,1]$, then, we transform C2.1 and C5.1 into the GP formulation based on proposition 1 and convert C6 via proposition 2.

**Proposition 1:** Assuming $t_1$ as the index of iterations in Step 1, we rewrite $\log_2(\frac{F_r}{N_r(t_1)} \gamma_{r,n}(t_1)) \approx \log_2(F_r \gamma_{r,n}(t_1)) - \log_2(N_r(t_1))$ and replace it in (5). Now, by using DC-approximation we can achieve linear approximation of $\log_2(N_r(t_1))$ as

$$\log_2(N_r(t_1)) \approx \log_2(N_r(t_1 - 1)) + \tag{7}$$
$$\nabla \log_2(N_r(t_1 - 1))(N_r(t_1) - N_r(t_1 - 1)),$$

where $N_r = \sum_{n \in \mathcal{N}} \alpha_{r,n}$. Further simplifying (7), we have

$$\log_2(N_r(t_1)) \approx \log(N_r(t_1 - 1)) + \tag{8}$$
$$\sum_{n \in \mathcal{N}} \frac{\alpha_{r,n}(t_1)}{\sum_{n \in \mathcal{N}} \alpha_{r,n}(t_1 - 1)} - \sum_{n \in \mathcal{N}} \frac{\alpha_{r,n}(t_1 - 1)}{\sum_{n \in N} \alpha_{r,n}(t_1 - 1)},$$

where via replacing (8) into C2, we have

$$\sum_{r \in \mathcal{R}} \alpha_{r,n}(t_1)[\log_2(F_r \gamma_{r,n}(t) - \log_2(N_r(t_1 - 1)) -$$

$$\sum_{n \in \mathcal{N}} \frac{\alpha_{r,n}(t_1)}{\sum_{n \in \mathcal{N}} \alpha_{r,n}(t_1-1)} + \sum_{n \in \mathcal{N}} \frac{\alpha_{r,n}(t_1-1)}{\sum_{n \in \mathcal{N}} \alpha_{r,n}(t_1-1)}] \geq R_n^{\text{rsv}}, \forall n \in \mathcal{N},$$

now, by applying AGMA, we reach to GP form of C2.1 as

$$\tilde{C}2.1: [R_n^{\text{rsv}} + \sum_{r \in \mathcal{R}} \alpha_{r,n}(t_1)[\log_2(N_r(t_1-1)) +$$

$$\sum_{n \in \mathcal{N}} \frac{\alpha_{r,n}(t_1)}{\sum_{n \in \mathcal{N}} \alpha_{r,n}(t_1)}]] \prod_{r \in \mathcal{R}} \left( \frac{\alpha_{r,n}(t_1)\log_2(F_r \gamma_{r,n}(t))}{\lambda_r(t_1)} \right)^{-\lambda_r(t_1)}$$

$$\prod_{m \in \mathcal{R}} \left( \frac{\alpha_{r,n}(t_1) \sum_{n \in N} \frac{\alpha_{r,n}(t_1-1)}{\sum_{n \in N} \alpha_{r,n}(t_1-1)}}{\phi_r(t_1)} \right)^{-\phi_r(t_1)} \leq 1,$$

Where

$$\lambda_r(t_1) = \qquad (9)$$

$$\frac{\alpha_{r,n}(t_1-1)\log_2(F_r \gamma_{r,n}(t))}{\sum_{r \in \mathcal{R}} \alpha_{r,n}(t_1-1) \left( \log_2(F_r \gamma_{r,n}(t)) + \sum_{n \in N} \frac{\alpha_{r,n}(t_1-1)}{\sum_{n \in N} \alpha_{r,n}(t_1-1)} \right)},$$

and

$$\phi_r(t_1) = \qquad (10)$$

$$\frac{\alpha_{r,n}(t_1-1) \sum_{n \in N} \frac{\alpha_{r,n}(t_1-1)}{\sum_{n \in N} \alpha_{r,n}(t_1-1)}}{\sum_{r \in \mathcal{R}} \alpha_{r,n}(t_1-1) \left( \log_2(F_r \gamma_{r,n}(t)) + \sum_{n \in N} \frac{\alpha_{r,n}(t_1-1)}{\sum_{n \in N} \alpha_{r,n}(t_1-1)} \right)}.$$

Similar to C2.1, DC-approximation is used to C5.1 and we can rewrite it as

$$\sum_{r \in \mathcal{R}} \sum_{n \in \mathcal{N}} \alpha_{r,n}(t_1) \beta_{r,n}(t_1)[\log_2(F_r \gamma_{r,n}(t) - \log_2(N_r(t_1-1))$$

$$-\sum_{n \in \mathcal{N}} \frac{\alpha_{r,n}(t_1)}{\sum_{n \in \mathcal{N}} \alpha_{r,b}(t_1-1)} + \sum_{n \in \mathcal{N}} \frac{\alpha_{r,n}(t_1-1)}{\sum_{n \in \mathcal{N}} \alpha_{r,n}(t_1-1)}] \leq L_b^{\max}, \forall b \in \mathcal{B}.$$

Now, by using AGMA, we have GP form of C5.1 as

$$\tilde{C}5.1:$$

$$\sum_{r \in \mathcal{R}} \sum_{n \in \mathcal{N}} \alpha_{r,n}(t_1) \beta_{r,n}(t_1)[\log_2(F_r \gamma_{r,n}(t) +$$

$$\sum_{n \in \mathcal{N}} \frac{\alpha_{r,n}(t_1-1)}{\sum_{n \in \mathcal{N}} \alpha_{r,n}(t_1-1)}] \left( \frac{L_b^{\max}}{\psi(t_1)} \right)^{-\psi(t_1)}$$

$$\prod_{r \in \mathcal{R}} \left( \frac{\alpha_{r,n}(t_1) \beta_{r,b}(t_1) \log_2(N_r(t_1-1))}{\xi(t_1)} \right)^{-\xi(t_1)}$$

$$\prod_{n \in \mathcal{N}} \left( \frac{\alpha_{r,n}(t_1) \beta_{r,b}(t_1) \frac{\alpha_{r,n}(t_1)}{\sum_{n \in N} \alpha_{r,n}(t_1-1)}}{\rho(t_1)} \right)^{-\rho(t_1)} \leq 1, \forall b \in \mathcal{B}$$

where

$$I(t_1) = L_b^{\max} + \sum_{r \in \mathcal{R}} \sum_{n \in \mathcal{N}} \alpha_{r,n}(t_1-1) \beta_{r,b}(t_1-1) \log_2(N_r(t_1-1) +$$

$$\sum_{r \in \mathcal{R}} \sum_{n \in \mathcal{N}} \alpha_{r,n}(t_1-1) \beta_{r,b}(t_1-1) \sum_{n \in N} \frac{\alpha_{r,n}(t_1-1)}{\sum_{n \in N} \alpha_{r,n}(t_1-1)}, \qquad (11)$$

$$\psi(t_1) = \frac{F_{r,b}^{\max}}{I(t_1)}, \qquad (12)$$

$$\xi(t_1) = \qquad (13)$$

$$\frac{\alpha_{r,n}(t_1-1) \beta_{r,b}(t_1-1) \log_2(N_r(t_1-1))}{I(t_1)},$$

$$\rho(t_1) = \frac{\alpha_{r,n}(t_1-1) \beta_{r,b}(t_1-1) \sum_{n \in N} \frac{\alpha_{r,n}(t_1-1)}{\sum_{n \in N} \alpha_{r,n}(t_1-1)}}{I(t_1)}. \qquad (14)$$

**Proposition 2:** Due to existence of negative terms in C6, it does not satisfy the conditions of posynomials in GP formulations. Therefore, by adding 1 to both the left and right hand sides of C6, we have $C6: \sum_{b \in \mathcal{B}} \beta_{r,b} + 1 \leq y_r + 1.$ Now, by applying AGMA technique, we achieve the monomial approximation for C6 as [23]

$$\tilde{C}6: (\sum_{b \in \mathcal{B}} \beta_{r,b} + 1) \times (\frac{1}{\mu(t_1)})^{-\mu(t_1)}$$

$$\prod_{r \in \mathcal{R}} \times (\frac{y_r(t_1)}{\varphi(t_1)})^{-\varphi(t_1)} \leq 1, \forall r \in \mathcal{R}$$

where

$$\mu(t_1) = \frac{1}{1 + y_r(t_1-1)}, \qquad (15)$$

and

$$\varphi(t_1) = \frac{y_r(t_1-1)}{1 + y_r(t_1-1)}. \qquad (16)$$

Consequently, at the iteration $t_1$, the GP approximation of (6) is

$$\min_{\boldsymbol{\alpha}, \boldsymbol{\beta}, \mathbf{Y}} U(\boldsymbol{\alpha}, \mathbf{P}(t), \mathbf{Y}), \qquad (17)$$

subject to: $\tilde{C}2.1, C3, C4, \tilde{C}5.1, \tilde{C}6, C7.$

Iteratively, the optimization problem (17) can be solved via on-line available soft wares such as CVX [29]. Step 1 will be stopped if the convergence criteria

$||\boldsymbol{\alpha}^*(t) - \boldsymbol{\alpha}^*(t-1)|| \leq \epsilon_1$, $||\boldsymbol{\beta}^*(t) - \boldsymbol{\beta}^*(t-1)|| \leq \epsilon_2$ and $||\mathbf{Y}^*(t) - \mathbf{Y}^*(t-1)|| \leq \epsilon_3$, are met and the optimal values of $\boldsymbol{\alpha}^*(t), \boldsymbol{\beta}^*(t)$ and $\mathbf{Y}^*(t)$ are achieved.

*Step 2: Power Allocation Algorithm*

For fixed values of $\boldsymbol{\alpha}^*(t), \boldsymbol{\beta}^*(t)$ and $\mathbf{Y}^*(t)$ achieved from Step 1, the optimization problem for power allocation in Step 2 is

$$\min_{\mathbf{P}} U(\mathbf{P}(t_2)), \tag{18}$$

subject to: C1,

$$C2.2: \sum_{r \in \mathcal{R}} \alpha_{r,n} \tilde{R}_{r,n}(\mathbf{P}(t_2)) \geq R_n^{\text{rsv}}, \forall n \in \mathcal{N},$$

$$C5.2: \sum_{r \in \mathcal{R}} \sum_{n \in \mathcal{N}} \beta_{r,b} \alpha_{r,n} \tilde{R}_{r,n}(\mathbf{P}(t_2)) \leq L_b^{\max}, \forall b \in \mathcal{B}$$

where the index of iterations in Step 2 is $t_2$. In (18), the only optimization variable is $\mathbf{P}$. Hence, (18) has less computational complexity compared to that (3). (18) is non-convex due to the non-linear logarithm terms in C2.2 and C5.2. To overcome computational complexity, at iteration $t_2$, we first apply DC approximation of $\tilde{R}_{k,s,n}(\mathbf{P})$ and then by applying AGMA, we will transform (18) into GP approximation as shown in the Proposition 3.

**Proposition 3:** We can rewrite C2.2 for user $n \in N$ as

$$\sum_{r \in \mathcal{R}} \alpha_{r,n}(t) \beta_{r,n}(t) \left( \log_2 \left( \frac{\frac{F_r}{N_r(t)} p_{r,n}(t_2) h_{r,n}}{\sigma^2 + I_{r,n}(t_2)} \right) \right) \geq R_n^{\text{rsv}},$$

which can be mathematically represented as

$$\log_2 \prod_{r \in \mathcal{R}} \alpha_{r,n}(t) \beta_{r,n}(t) \left( \frac{\frac{F_r}{N_r(t)} p_{r,n}(t_2) h_{r,n}}{\sigma^2 + I_{r,n}(t_2)} \right) \geq R_n^{\text{rsv}},$$

therefore, we reach to

$$\tilde{C}2.2: \prod_{\forall r \in \mathcal{R}} \alpha_{r,n}(t) \left( \frac{\sigma^2 + I_{r,n}(t_2)}{\frac{F_r}{N_r(t)} p_{r,n}(t_2) h_{r,n}} \right) \leq 2^{-R_n^{\text{rsv}}}.$$

In order to decrement of computational complexity, we introduce the predefined threshold for tolerate interference as $I_{r,n}^{\text{th}}$ and it used in C5.2 instead of $I_{r,n}^{\text{th}}$ which leads to the less limitation of network rate [31-33]. By this assumption C5.2 is converted into GP standard form as

$$\tilde{C}5.2: \prod_{\substack{r \in \mathcal{R} \\ n \in \mathcal{N}}} \beta_{r,b}(t) \alpha_{r,n}(t) \left( \frac{\frac{F_r}{N_r(t)} p_{r,n}(t_2) h_{r,n}}{\sigma^2 + I_{r,n}^{th}} \right)$$

$$\leq 2^{L_b^{\max}}, \forall b \in \mathcal{B}.$$

Consequently, at the iteration $t_2$, the GP approximation of (15) is

$$\min_{\mathbf{p}} U(\mathbf{p}(t_2)), \tag{19}$$

subject to: C1, $\tilde{C}$2.2, $\tilde{C}$5.2.

Iteratively, (19) is solved until the convergence criteria $||\mathbf{P}^*t_2 - \mathbf{P}^*(t_2 - 1)|| \leq \epsilon_4$, are met. Since the proposed algorithm is a kind of the block SCA method, its convergence is guaranteed [22, 34, 35].

## IV. SIMULATION RESULTS

To study the performance of our approach, we consider that users are randomly located within the area served by $R = 5$ RRHs and $B = 2$ BBUs. The channel power loss between user $n \in \mathcal{N}$ located at a distance $d_{r,n} > 0$ from RRH $r \in \mathcal{R}$, is modeled as $h_{r,n} = \frac{1}{1 + (d_{r,n})^4}$. The values of maximum BBU load and the number of antennas mounted on the RRH $r \in \mathcal{R}$ are randomly chosen in the range of $L_b^{\max} \in [2, 24]$ and $F_r \in [100, 200]$, respectively. Furthermore, we set $C_a = 0.25$, $\epsilon_1 = \epsilon_2 = \epsilon_3 = \epsilon_4 = 10^{-3}$ and $p_r^{\max} = 40$ Watt ($\forall r \in \mathcal{R}$) for all of the computations.

In order to evaluate the efficiency of our proposed algorithm, we compare it with the traditional algorithm of wireless network. Hence, we choose the max SINR approach as traditional algorithm for user association. In traditional wireless networks, each user is assigned to the RRH based on the largest average received SINR [8], [36] as described by

Associated user n to RRH r =

$\text{argmax}_r \{SINR_r\}, \forall r \in \mathcal{R}, \forall n \in \mathcal{N},$

meaning that, based on reference signal received power (RSRP) broadcasted by RRHs [37], each user $n \in \mathcal{N}$ calculates received SINR from all RRHs and connects to the RRH with the largest received SINR. SINR for user $n \in \mathcal{N}$ at RRH $r \in \mathcal{R}$ is calculated as $\frac{p_{r,n} h_{r,n}}{\sigma^2 + I_{r,n}}$, in which $I_{r,n} = \sum_{\forall r' \in \mathcal{R}, r' \neq r} \sum_{\forall n' \neq n} p_{r',n'} h_{r,n'}$ is the interference to user $n \in \mathcal{N}$ at RRH $r \in \mathcal{R}$. Therefore, user association to RRHs is fixed and predetermined. In traditional algorithm, we consider all of the RRHs are in state on and they cannot be switched off (i.e., $y_r = 1, \forall r \in \mathcal{R}$). Consequently, the resource allocation problem (3) can be formulated as

$$\min_{\boldsymbol{\beta},\mathbf{P}} U(\mathbf{P}), \qquad (20)$$

subject to: $C1-C5$.

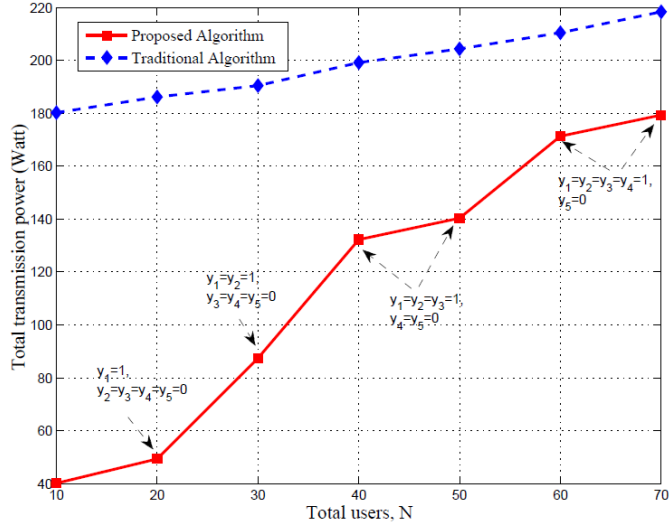

Fig. 2. Total transmission power versus N with $F_1 = F_2 = F_3 = F_4 = F_5 = 140$ and $R_n^{rsv} = 0.3 (bps/Hz)$.

In (20) the only optimization variables are $\boldsymbol{\beta}$ and $\mathbf{P}$. Therefore, similar to (3), we decompose (20) into two sub problems and apply CGP to solve it. Thus, with a given (fixed) user allocation vector (i.e., $\boldsymbol{\alpha}$), fronthaul links and RRHs are assigned to BBUs and power is allocated to each user. Also, in the traditional algorithm, the antennas of each RRH are equally divided between the users that connected to it.

The aim of purposed approach is to minimize the total power consumption cost by switching off the under-utilized RRHs. The RRHs with low data traffic are switched off and users assigned to them are moved to the neighboring RRHs. Therefore, the number of active RRHs and power consumption cost of network will be minimized. Compared to that, in traditional scenario, users are assigned to the RRHs regardless of minimizing the overall network power, hence, none of the RRHs will be switched off and user association is predefined.

In Fig. 2, the total transmission power versus the number of users is illustrated for both our proposed approach and the traditional scenario. From Fig. 2, it is clear that the total transmission power increases with increasing $N$ for both cases. However, the total transmission power in our proposed algorithm is less than that of traditional scenario. This is because, in order to minimize the power consumption cost of network, RRH association manages the interference between RRHs which leads to switch off under-utilized RRHs while the RRH assignment to each user is predefined in traditional scenario. For instance, in our proposed algorithm, when data traffic is low, e.g., $N = 20$, Fig. 2 illustrates that with only one active RRH (e.g., $y_1=1, y_2=y_3=y_4=y_5 0$) the minimum required rate of users will be satisfied while all five RRHs are in state on for traditional algorithm. Reducing the number of active RRHs can lead to power saving of the RRHs. Also, Fig. 2 shows that with increasing traffic demand, we need the more number of active RRHs to satisfy the $R_n^{rsv}$ of users. For instance, in proposed algorithm, when $N = 70$ four RRHs are activated and the total transmission power is close to the traditional algorithm.

Besides, Fig. 2 indicates that with increasing $N$, the total transmission power is linearly increased via proposed approach. Note that, for $N \in [40, 50]$, this transmission power with increasing $N$ does not have considerable increment while by switching on a new RRH, there will be a sudden increase in the total transmission power. For instance, when $N = 50$, three RRHs are in state on, but when $N = 60$, in addition to the previous active RRHs, RRH 4 should be switched on $(y_4 = 1)$.

In Fig.3, the effect of minimum required rate of each user, e.g., $R_n^{rsv}$, on the total transmission power is demonstrated. Fig.3 indicates that with increasing $R_n^{rsv}$, to meet C2, the total transmission power is increased. This is because, with increasing $R_n^{rsv}$ and traffic demand, the feasibility regions of resource allocation hold which leads to switch on the more number of RRHs and increase power consumption cost of RRHs. Fig.4 indicates the total throughput versus the total number of users for both algorithms. Based on multiuser diversity gain [38], Fig.4 demonstrates that the total throughput is increased with increasing the number of users for both algorithms. Fig.4 shows when data traffic is low (N=20) in our proposed algorithm, the total throughput is less than that of traditional algorithm. This is because in our proposed algorithm only one RRH is in state on but in traditional algorithm all of the RRHs are in state on. For this reason, the feasibility region of resource allocation in traditional algorithm is larger than that of our proposed algorithm, therefore, users are closer to the RRHs and they can get larger SINR with less interference and transmit power compare to that of our proposed algorithm. Besides, Fig. 4 shows that with increasing traffic demand (after N=40), the total throughput in our proposed algorithm, will be higher than the traditional scenario. It is mainly because, with increasing data traffic the more number of RRHs are switched on via our proposed algorithm, but their number is still lower than traditional scenario. Hence, our proposed algorithm with the less number of active RRHs can considerably control interference between RRHs and it can provide better coverage of users for the dense network.

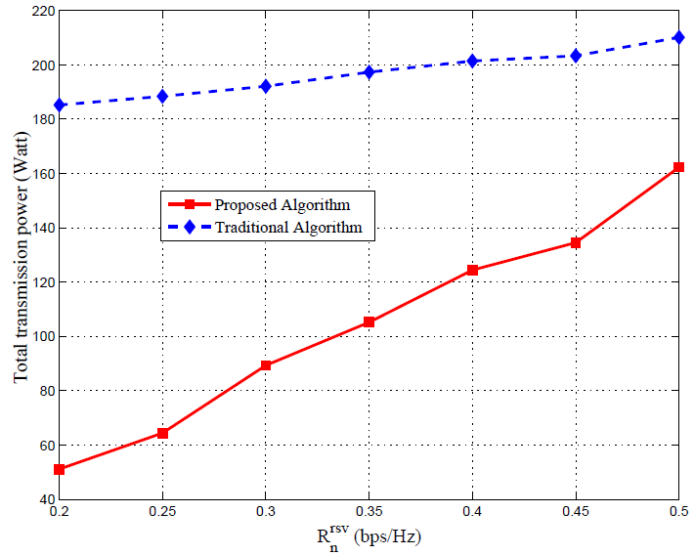

Fig. 3. Total transmission power versus $R_n^{rsv}$ with $F_1 = F_2 = F_3 = F_4 = F_5 = 140$ and $N = 30$.

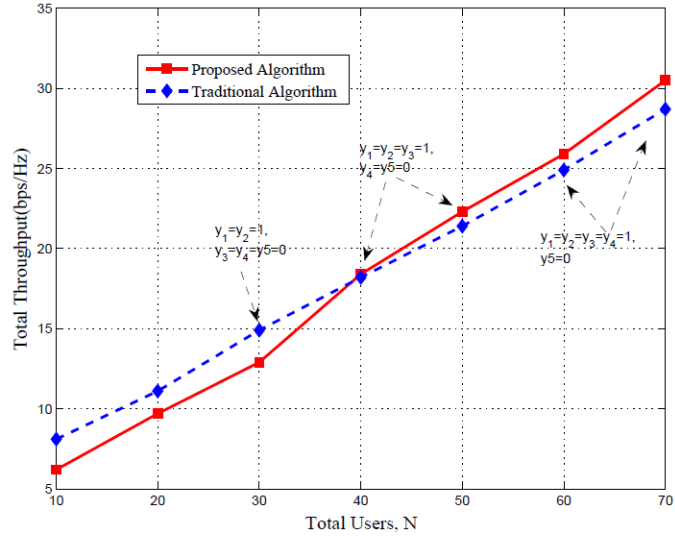

Fig. 4. Total throughput versus N with $F_1 = F_2 = F_3 = F_4 = F_5 = 140$ and $R_n^{rsv} = 0.3 (bps/Hz)$.

Therefore, the total achieved throughput will be better than traditional algorithm. Note that, in high data traffic, providing higher throughput is more significant than decreasing power consumption cost, therefore by switching on the more number of RRHs in a cell, spectrum efficiency and the total throughput of the network will be improved although overall network power will be increased.

Based on our last knowledge, the majority of pervious works, e.g., [15], [39], EE has been defined as the ratio of the achievable sum rate and the sum power consumption as

$$EE = \frac{\sum_{n \in \mathcal{N}} R_n}{P_{total}}, \quad (21)$$

which $\sum_{n \in \mathcal{N}} R_n$ and $P_{total}$ are total throughput and total transmit power, respectively. Hence, we consider

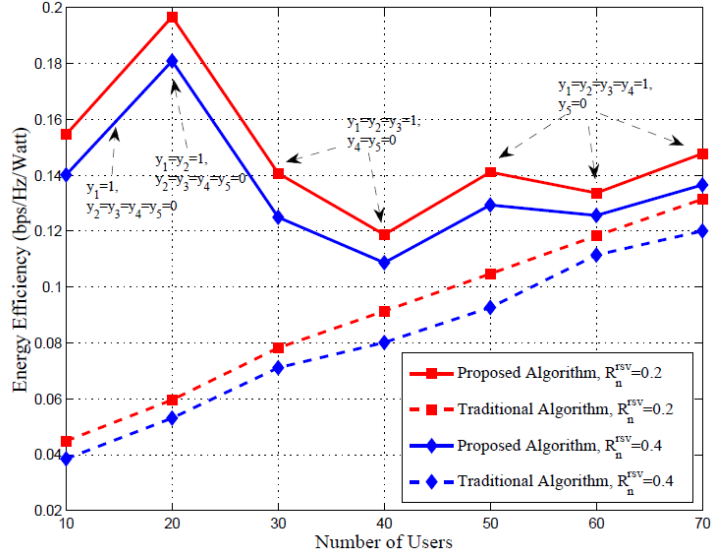

Fig. 5. Energy efficiency versus N with $F_1 = F_2 = F_3 = F_4 = F_5 = 140$

$P_{total}$ as the total energy consumption cost of network which is defined in utility function (2). Therefore, based on (21) and (2), we can define EE as the ratio of the total achieved throughput to the total energy consumption cost of network in units of bps/Hz/Watt, which is given by

$$EE(\boldsymbol{\alpha},\mathbf{P},\mathbf{Y}) = \frac{\sum_{r \in R}\sum_{n \in N} \alpha_{r,n} R_{r,n}(\mathbf{P},\boldsymbol{\alpha})}{\sum_{r \in R}\sum_{n \in N} \alpha_{r,n} p_{r,n} + C_a y_r F_r}. \quad (22)$$

In Fig. 5, we show the EE versus the total number of users for our proposed approach and traditional scenario. Fig. 5 demonstrates that EE increases by increasing the number of users in traditional algorithm. Also, the simulation results illustrate that via proposed approach with increasing the number of users, EE increases, but as soon as a new RRH is switched on, EE will be reduced. This is because with increasing traffic demand, the more number of RRHs should be switched on to satisfy $R_n^{rsv}$ of users, which leads to the more power consumption cost and decrement of EE. For instance, Fig. 5 indicates that immediately after $N=20, N=30$ and N=50, RRHs 2, 3, and 4 are switched on, respectively. Therefore, the total transmission power is linearly increased and EE will be decreased. Moreover, from the simulation results it can be seen that in both algorithms, with increasing the value of $R_n^{rsv}$, the EE decreases because the feasibility region of resource allocation in (3) is decreased which leads to less total achieved throughput. However, due to interference management and RRH association to each user via proposed approach, the chance to allocate feasible transmit power between users will be increased which can provide better EE compared to that of traditional Algorithm.

## V. CONCLUSION

In this paper, we study power efficient optimization problem in massive MIMO aided C-RANs, where the RRHs can be switched off to save energy. We formulate this problem as a joint RRH and power allocation to each user, and fronthaul link and RRH-BBU assignment optimization problem while the minimum required rate of each user should be obtained. To minimize the total network power consumption, we develop an efficient two-step iterative algorithm to dynamically allocate resources. The simulation results illustrate that our proposed scheme is more efficient compared to the traditional algorithm, in terms of minimizing overall network transmission power which leads to improve EE.